\documentclass[a4paper, 12pt]{article}
\usepackage{chicago}
\usepackage[pdftex]{graphicx}
\title{Consensus, Cohesion and Connectivity}
\author{Jeroen Bruggeman\thanks{Department of Sociology, University of Amsterdam, Nieuwe Achtergracht 166, 1018 WV Amsterdam, the Netherlands. Email: j.p.bruggeman@uva.nl. }}
\date{\today}
\begin{document}  
\maketitle

\begin{abstract}  
Social life clusters into groups held together by ties that also transmit information. When collective problems occur, group members use their ties to discuss what to do and to establish an agreement, to be reached quick enough to prevent discounting the value of the group decision. The speed at which a group reaches consensus can be predicted by the algebraic connectivity of the network, which also imposes a lower bound on the group's cohesion. This specific measure of connectivity is put to the test by re-using experimental data, which confirm the prediction.   
\end{abstract}

\section{Introduction}

Social life coalesces into groups that face various problems, including dilemma's of collective action. Many of these problems involve uncertainties due to social interdependencies. Then, group members can not simply decide on the basis of expected costs and benefits; people may disagree on what the problem is, have different views on solving it, wonder which hurdles will be in the way, question how the contributions are to be divided, along skills or other traits, and doubt to what extent they will benefit from the, currently fuzzy, solution. Throughout history, from hunter-gatherers to teams and committees, people have dealt with such ill structured problems \cite{simon73}, and most of the time without centralized leadership \cite{gintis15}. Examples are food shortages, threats by competing groups, environmental pollution or the development of new norms. To reach a shared intentionality \cite{tomasello05}, or jointness \cite{lawler01}, with respect to a collective problem and its solution, people will discuss the matter. Because people discount the value of future outcomes \cite{levin12}, however, an agreement, if reached at all, is to be reached relatively quickly. 

Both consensus and the speed at which it is reached depend on the topology of participants' network, which has been examined at book length by Friedkin and Johnsen (2011), building on Festinger \cite {festinger50}, French \cite{french56} and other social psychologists. I contribute to their theory of social influence by showing that the speed of reaching a consensus in non-centralized group decisions is determined by a single parameter of the network, the so called \textit{algebraic connectivity} \cite{fiedler73}. Networks with high algebraic connectivity appear roundish without large holes in them, whereas networks with low connectivity have skinny parts, large distances or a highly heterogeneous degree distribution. Algebraic connectivity increases the predictive power of social influence theory, shown by re-using experimental data (Section~3). Moreover, it makes possible a clear-cut relation with another key concept in the social sciences---\textit{social cohesion} (Section~4)---that was defined considering a multiplicity of independent connections \cite{moodywhite03}. In the final section (5), the results are summarized, and limitations of algebraic connectivity are indicated. The next section introduces the Friedkin-Johnsen model and Laplacian matrices, which are stepping stones towards this paper's results.

\section{Social influence}
Suppose that $n$ people gather at a \textit{social focus} \cite{feld81}, in our case a problem that affects all of them. Finding themselves in the same situation under uncertainty will make them at least somewhat receptive to one another's ideas. They will start discussing the problem or its symptoms, possible solutions, moral justifications, obstacles as well as contributions to be made. They may already know one another and communicate on a prior-established network, but if not, their communications will establish a network in situ. I do not assume that all participants have the same commitment to finding a solution, although all of them are supposed to have some. 

Experiments show that if two people communicate (and do not strongly dislike each other in the first place), ego's and alter's ideas encroach upon each other gradually \cite{friedkin11}. This empirical regularity, replicated several times experimentally \cite{takacs16,moussaid15,kerckhove16}, is the micro foundation of social influence theory.\footnote{I do not assume that in every social situation, people will reach an agreement; see for example \cite{friedkin16,aral12}.}   

For simplicity, I will focus on one issue under discussion, $y$, rather than a multitude of them, but the social influence model can be generalized to multiple issues \cite{friedkin11}. Individuals $i$ and $j$ reach consensus when their opinions, or expectations, are very similar up to some (small) level of indifference, $|y_j - y_i| < \epsilon$. This definition can be easily generalized to all pairwise differences in a group.

Participants' communication pattern establishes, or uses, a social network, represented by an adjacency matrix $\bf{A}$. For consensus to be reached, the network should be one component, where everybody is connected indirectly or directly to everybody else. Although we start out with binary, symmetric ties, we have to take into account that when someone is simultaneously influenced by multiple alters, the influence of an individual contact will weaken amidst a larger number of them. Therefore the adjacency matrix is first row-normalized, denoted $\bf{W}$; the cell entries $w_{ij}$ of each row $i$ sum to one. (There is in fact also a technical reason for row-normalization, which is not important for the current discussion.) Friedkin and Johnsen \cite[p.14]{friedkin11} model the change of focal actor $i$'s opinion $y_i$ from $t = 1$ to $t = 2$ in response to communication with social contacts $j$ as:

\begin{equation}
y_i^{(2)} - y_i^{(1)} = s_{ii} \left(\sum_{j=1}^{n} w_{ij}y_j^{(1)} - y_i^{(1)}\right).
\label{eq:influence1}
\end{equation} 
In this model, $s_{ii}$ ($0 \leq s_{ii} \leq 1$) denotes $i$'s influenceability, or susceptibility to social influence from other group members, which turned out to be in the range 0.3 to 0.5 under experimental conditions \cite{moussaid15,moussaid13,soll09}.  In all likelihood, susceptibility will increase with solidarity,\footnote{Solidarity is ``an identification with a collectivity such that an individual feels as if a common cause and fate are shared" \cite{hunt04}.  Whereas solidarity is someone's bonding to an entire group, cohesion is based on bonds between specific individuals that establishes the topology of the network (Section~4).\label{note:solidarity}} commitment to, or identification with the group or it's current goal \cite{klandermans02}.  The lower $i$'s solidarity or commitment, the less she will let herself be influenced, and the smaller the difference between $y_i^{(2)}$ and $y_i^{(1)}$ will be. 

We now proceed beyond the Friedkin-Johnsen model by examining continuous change of opinion \cite{abelson64}, denoted $dy_i/dt$, rather than change between discrete time points. For the moment, we leave out susceptibility, assuming that everybody in the group has roughly the same level of it. Using matrix notation, change of opinion can now be described very parsimoniously for all group members in one stroke:

\begin{equation}
\frac{d\bf{y}}{dt} = - \bf{L}\bf{y}.  
\label{eq:influence2}
\end{equation}  
In this model, $\bf{L}$ is the Laplacian matrix, which is an alternative representation of a graph that contains the same information as the adjacency matrix. Analogous to the latter, the Laplacian has both a standard and a normalized form (and then some more). The standard Laplacian applies to graphs with symmetric, binary ties; its cells are obtained from the adjacency matrix by making them negative (ties then get the value -1) and keeping the zeros for unconnected nodes. Subsequently, the degrees of the nodes are put at the diagonal. In matrix notation, the adjacency matrix is subtracted from a diagonal matrix, $\bf{D}$ = $diag(\sum_j a_{ij}) $, that has zeros everywhere else,  
\begin{equation}
\bf{L} = \bf{D} - \bf{A}.
\label{eq:laplace1}
\end{equation}  
This definition makes it possible to relate opinion dynamics to social cohesion (Section~4). To stay close to the Friedkin-Johnsen model, however, the normalized Laplacian \cite{chung97} is used in the next section on experimental networks. In matrix notation, with $\bf{L}$ and $ \bf{D}$ as above, the normalized Laplacian is defined as:
\begin{equation}
\bf{\mathcal{L}} = \bf{D}^{-1/2} \bf{L} \bf{D}^{-1/2}.
\label{eq:laplace2}
\end{equation}  
When using  $\bf{\mathcal{L}}$ at the right hand side of Equation~\ref{eq:influence2} and performing the multiplication (and rearranging terms), it becomes clear that opinion dynamics depends on the dyadic differences of opinions, as in the Friedkin-Johnsen model. Both models show for a given $n$ and topology that the higher the diversity of initial opinions, the longer it will take for consensus to be reached. The two models are very similar, and if all nodes have the same degree, for instance in a clique, they are identical, be it that the Laplacian version makes it possible to make new predictions.  

Both the standard and the normalized Laplacian have a spectrum of positive eigenvalues, $0 = \lambda_1 \leq \lambda_{2} \leq \cdots \leq \lambda_N$; if there are $m$ (near) zero eigenvalues they indicate the presence of $m$ (almost) disconnected graph components that won't reach consensus with each other. For given initial opinions, consensus is reached faster when $\lambda_{2}$, called the \textit{algebraic connectivity} of the graph, is higher \cite{olfati04}.  Consensus speed can't be inferred from familiar network measures such as clustering, density, average path distance, degree distribution, degree centralization, size, $k$-core,\footnote{In a $k$-core, everybody has at least $k$ ties with others who in turn have at least $k$ ties \cite{seidman83}.} $\kappa$-connectivity (Section~4) or an eigenvalue of the adjacency matrix, even though density and short average distance are correlated with algebraic connectivity.  Using a normalized Laplacian, algebraic connectivity's maximal value equals 2, for a dyad, equals 1.5 for a triad, and further decreases to 1 for large cliques. Numerically solving the model, $d{\bf y}/{dt} = - \bf{\mathcal{L}} \bf{y} $, for a given set of initial opinions in small cliques confirms that relatively larger cliques take more time to synchronize. 
 
\begin{figure}
\begin{center}
\includegraphics[width=.7\textwidth]{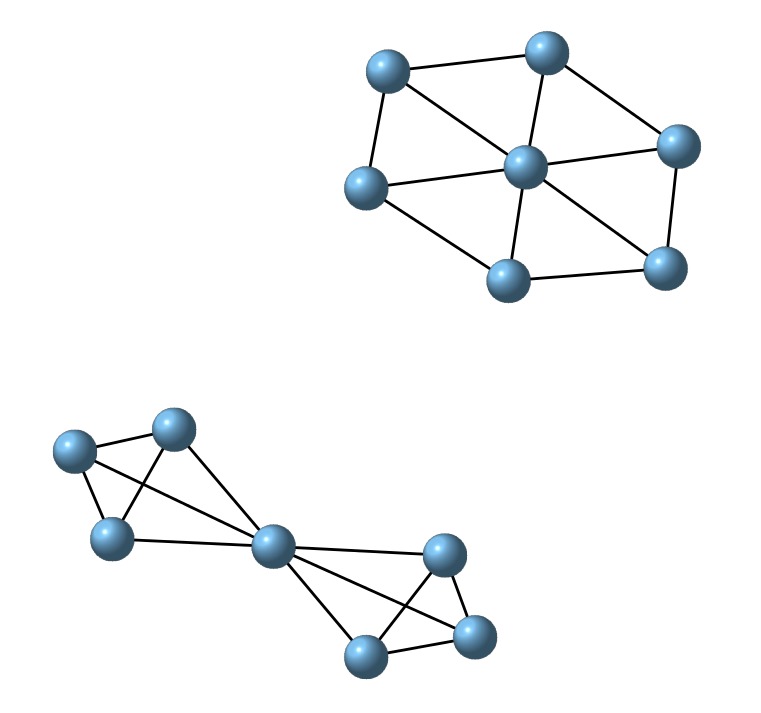}
\end{center}
\caption{Wheel and bow-tie networks.} 
\end{figure}
To illustrate, Figure~1 shows two networks with the same size (7), density (0.57), average  distance (1.43), degree distribution, degree centralization (0.6), and coreness (both are 3-cores). The bow-tie is more clustered (0.73) than the wheel (0.55), but in the wheel, with $\lambda_{2}=2/3$, consensus is predicted to be reached faster than in the bow-tie, with $\lambda_{2}=1/3$, confirmed by numerically solving the model for these two networks.  

Beyond this paper, we want a general model that is applicable to networks with variation of susceptibility across individuals, as well as weighted, possibly strongly asymmetric, ties.  To this end we can calculate a Laplacian defined for matrices with weighted asymmetric positive ties \cite{chung05}, % s_ii moet nog op de diagonaal van La ?
\begin{equation}
{\bf \mathcal{L}}_a = {\bf I} - \frac{ \Phi^{1/2} {\bf W} \Phi^{-1/2} + \Phi^{-1/2} {\bf W}^T \Phi^{1/2} }{2},
\label{eq:laplace3}
\end{equation}  
% compared with clique with binary ties, lambda_2 should be lower if tie weights are not equally distributed, and much lower (compared with single tie effect) if an individual has a lower a with the group as a whole.
where $\bf{I}$ is the identity matrix, and $\Phi$ the Perron vector of ${\bf W}$ written as a diagonal matrix.  For cliques, each cell in the Perron vector equals $1/n$, but in general, these cells can be estimated with PageRank \cite{prystowsky05}.  Again, the algebraic connectivity indicates the speed of convergence, and when the graph is binary and symmetric and the susceptibilities equal one, this Laplacian becomes identical to the normalized Laplacian above. The general influence model is: 
\begin{equation}
\frac{d\bf{y}}{dt} = - \bf{S}{\bf \mathcal{L}}_a\bf{y}, 
\label{eq:laplace4}
\end{equation}  
with individuals' susceptibilities written as a diagonal matrix, $\bf{S}$ = $diag(s_{ii})$. For this model (Eq.\ref{eq:laplace4}) it is no longer possible to assess the speed of convergence by algebraic connectivity,\footnote{The claim of the opposite, in the {\em Social Networks} version of this paper, is wrong.} but we can run the model on a computer. It shows, for a given topology, that if few people have relatively low susceptibility, they slow down the consensus of the entire group. Empirical applications of this model are left for future studies. Other researchers studied experimentally how susceptibility changes longitudinally \cite{friedkin16b}, and found that the five major personality factors are unrelated to it \cite{kerckhove16}. An interesting field study during the Libyan civil war (2011) pointed out that when people share strong negative emotions, in this case in combat, they more strongly identify with one another \cite{whitehouse14b}, which will increase their mutual influenceability. 

\section{Empirical evidence}
Evidence of algebraic connectivity speeding up consensus can be gleaned from the following experiment \cite{judd10}. Subjects ($n=36$) had to choose the same color (from a given set) as their network-neighbors to earn money. This is a local coordination problem, just like mounting a collective action without centralized leadership. The ties (symmetric and binary) in a highly clustered network were randomly rewired while keeping the clusters connected. Figure~2 at the top shows the clustered network and the resulting network when each tie has a probability of $p=0.2$ for one of its ends to be connected randomly to another node in the network. Probability values $p \in \{0, 0.1, 0.2, 0.4, 0.6, 1 \}$ yield six different networks with the same size and density. The mean time to reach consensus in each of these networks in the experiment is depicted at the bottom of Figure~2 (black dots). 
\begin{figure}
\begin{center}
%\begin{array}{c}
\includegraphics[width=.9\textwidth]{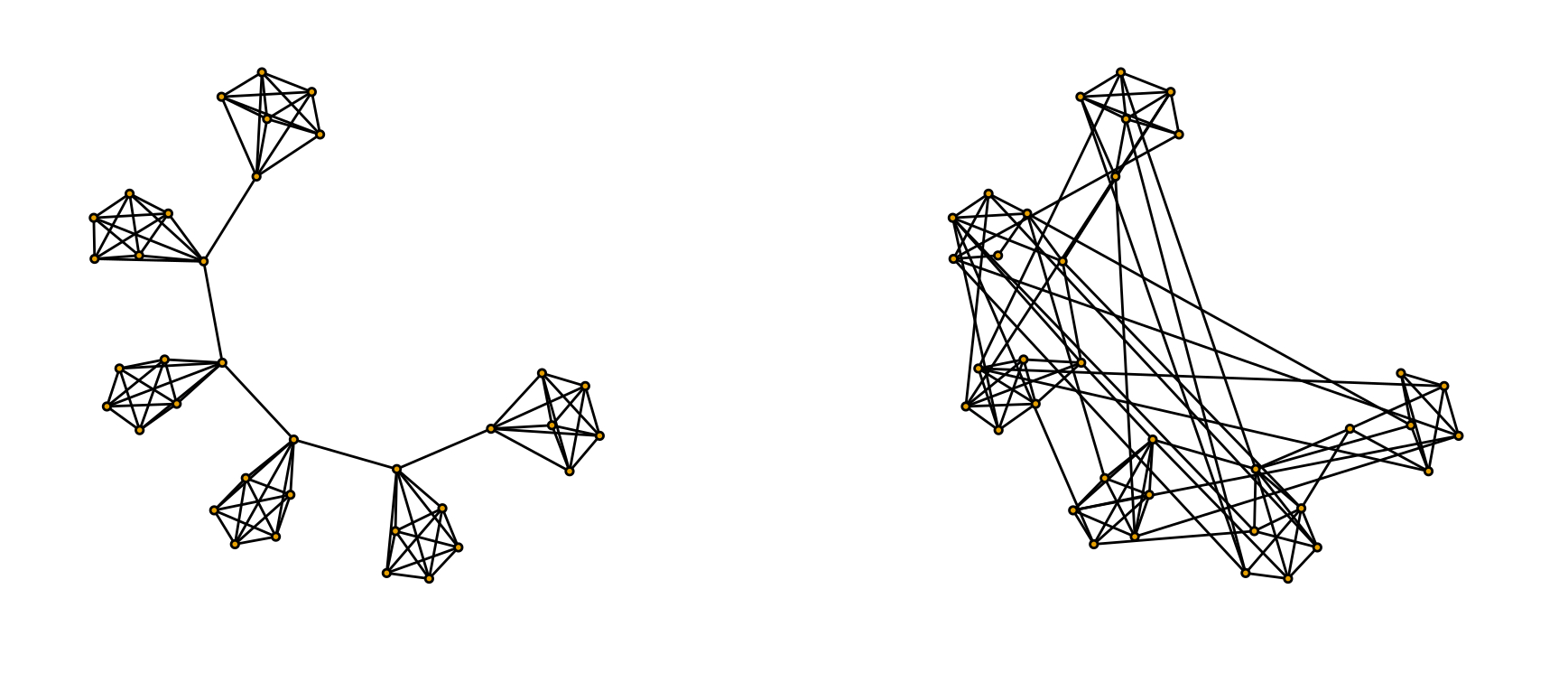}
\includegraphics[width=.9\textwidth]{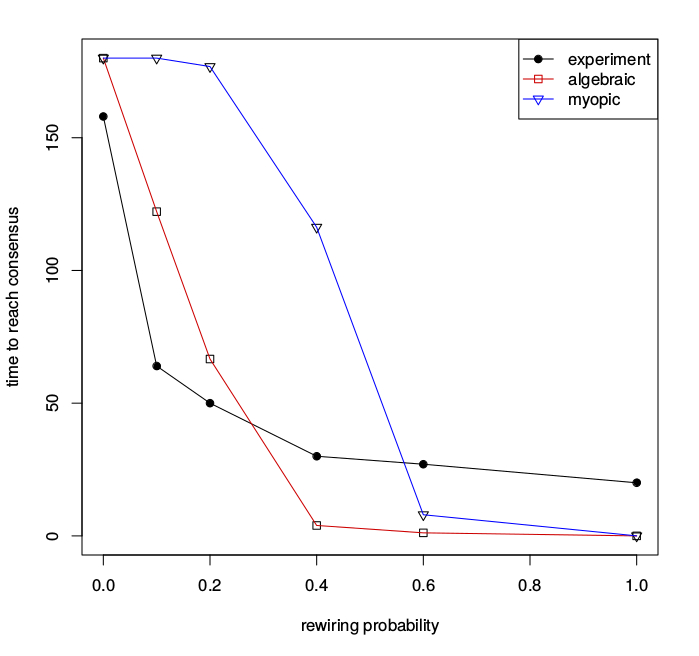}
%\end{array}
\end{center}
\caption{Graph coloring experiment. Top: Clustered network ($p = 0$) at the left and a randomly rewired network ($p = 0.2$) at the right.  Bottom: Subjects' time to reach consensus (black dots), algebraic connectivity (open squares) and the myopic conformity model (open triangles).} 
\end{figure}

We can approximate the discrete colors the subjects could choose from as a continuous variable, i.e.~wavelength of light, and apply the social influence model. Because everybody received the same amount of money when coordination succeeded, large differences of susceptibility are not to be expected.  For each probability larger than 0, I randomly rewired the clustered network a thousand times, and calculated the mean algebraic connectivity 
(0.0083; 0.104; 0.196; 0.300; 0.304; 0.306) of the normalized Laplacians (Eq~\ref{eq:laplace2}). Figure~2 shows that the increments of these values predict the decreasing times to reach consensus fairly well (open squares). The pertaining experimental study features a myopic conformity model that fits less well (open triangles).\footnote{Also DeGroot's model \cite{degroot74}, elaborated by Jackson \cite{jackson08}, explains consensus in this data, with a curve in between the myopic conformity model and algebraic connectivity. In contrast to the latter, it has no established relation with cohesion, though.}   
%\footnote{I asked the corresponding author for the original data, to which he did not reply. Then I enlarged the original figure (here Figure~2) and used a good old ruler to assess the consensus times. Simulating inaccuracy of assessment sometimes yielded a 0.01 difference with the numbers reported here.} 
Interestingly, the largest improvement of consensus time is had at small increases of randomness in highly clustered networks, just as the largest decrease of average distance in the small-world model \cite{watts98}.  

In theory, algebraic connectivity of sparse networks can be further increased by adding ties, but in actuality there are cognitive limitations to tie maintenance \cite{miritello13,saramaki14}. Moreover the experimental data suggest that not a great deal of further improvement of consensus time is to be expected.

Algebraic connectivity is also consistent with the result of a different experiment, on the creation of a group convention through local interactions \cite{centola15}.  In contrast to the graph coloring experiment, the subjects in the convention experiment interacted with only one alter in each round, over 30 rounds in total. Three different topologies, each with $n = 24$, were compared: a one-dimensional lattice with a degree of 4 ($\lambda_2 = 0.084$), a random network with an average degree of 4 ($\lambda_2 = 0.205$), and a homogeneous network where subjects were connected to a new contact in every subsequent round. This procedure implies that at round 23, the homogeneous network had developed into a clique ($\lambda_2 = 1.043$), when compressing all ties into one graph.  In the lattice and random networks, a group convention was never achieved during 30 rounds, in contrast to the homogeneous network, where it always was. Also in larger and sparser homogeneous networks ($n = 48; n = 96$), global conventions were achieved, even though the diversity of convention proposals increased with the number of subjects \cite{centola15}. These larger networks still have high algebraic connectivity; for a homogeneous network with $n = 96$ and a degree of, say, 20 (at round 20), $\lambda_2 = 0.63$, whereas the same sized random network with an average degree of 4 has only $\lambda_2 = 0.028$, and the lattice lags behind with $\lambda_2 = 0.0053$.  Although the model can not predict the round when a local convention becomes global, algebraic connectivity is consistent with the experimental result. This is remarkable, because the influence model was initially developed for stable networks with quantitative differences of opinions, not for cases where ties are toggled and opinions are categorical, as in this experiment. 

\section{Cohesion}
In the graph coloring and convention experiments, information that individuals received from their network neighbors was perfectly accurate. In actuality, however, there is \textit{noise}---misinterpreted, wrongly transmitted or strategically manipulated information. When information is transmitted through a single chain it deteriorates in every subsequent step \cite{moussaid15}. Transmission through two node-independent channels provides a substantial improvement, and the difference with a single channel becomes more noticeable over longer distances \cite{eriksson12}. More than two-channel transmission was not tested experimentally, but it's likely that the marginal improvement will decrease with the number of independent channels \cite[p.314]{whiteharary01}, and that two might be enough in daily social life.\footnote{If there is a great deal of noise, however, high algebraic connectivity won't save the day, and the group will turn into an echo chamber of (mostly) false information.}  %Another way to arrive at this conclusion is to see collecting information about or from someone as a parameter estimate, which typically improves with the square root of the number of observations (or ties) while the costs (of tie maintenance) increase linearly.  

When researchers conceptualized the notion of social cohesion, they realized that along with binding a group together, cohesion should also protect information transmission against noise \cite{whiteharary01,moodywhite03}. In this spirit, they defined the level of cohesion as the minimum number, $\kappa$, of node-independent channels connecting arbitrary pairs of nodes in a network (see footnote \ref{note:solidarity}). The notion of $\kappa$-cohesion thus implies that the ties that hold a group together also form redundant conduits of information.    

Because in general there is no relation between tie strength and reliability of transmitted information, I use the standard Laplacian for binary undirected ties (Eq.~\ref{eq:laplace1}).  
Now a mathematical theorem points out that for all incomplete networks (missing at least one tie), $\lambda_{2} \leq \kappa$ \cite{fiedler73,abreu07}. 
In the bow-tie in Figure~1, the left hand and right hand clusters receive news about each other only through the middle man ($\lambda_2 = 1; \kappa = 1$), whereas in the wheel, everybody is connected to everybody else through three independent channels ($\lambda_2 = 2; \kappa = 3$). Algebraic connectivity thus implicates a minimum number of node-independent information channels. 

\section{Discussion}
 
We have seen that algebraic connectivity can capture the combined effect of topology and tie strengths on the speed towards consensus, and that it implies a minimum level of social cohesion, which in turn makes a group more resilient against noise. When put to the test, the Laplacian version of the social influence model turns out to hold true even for categorical opinions and toggling ties, for which the model was initially not designed.

Seen from a broader perspective, we can see groups passing through different stages \cite{tuckman65}, while group members switch between different activities such as exploration and exploitation. For teams in modern organizations, for example, it was shown that clustering with cleavages between the clusters fosters exploration of new information. To turn this new information into innovations, in contrast, a more homogeneous topology is superior \cite{shore15}, but not with maximal connectivity. In this broader context, algebraic connectivity is a specific measure that predicts consensus, but it does not predict other things that people do collectively, such as establishing complementary roles \cite{judd10}, or exploring a changing environment and innovate to adapt to it. 

The applicability of algebraic connectivity is also limited for networks with centralized (and possibly formal) leadership. Patrons \cite{martin09} and managers \cite{simon97} have incentives at their disposal that enhance their influence beyond the pairwise differences of opinions in the model. Nevertheless, the model can handle differences of individuals' informal power due to their network positions, which can be analyzed through measures of power centrality \cite{bonacich87,friedkin91}.  If along with positive ties there are also negative ties, yet another Laplacian can be applied, developed for signed graphs \cite{kunegis10}. In connected signed networks, the smallest eigenvalue of this Laplacian indicates how unbalanced the network is \cite{facchetti11}. 

The main result of this study is that in groups without centralized leadership, consensus is achieved more quickly if participants' network has higher algebraic connectivity. Such networks existed from ancestral foragers to modern teams, and algebraic connectivity also guarantees a minimum level of social cohesion of these groups.

\end{document}